\documentclass[aps,prl,twocolumn,superscriptaddress,floatfix,showpacs]{revtex4}
\usepackage{graphicx}
\begin{document}

\title{Graphite intercalation compounds under pressure}

\author{G\'abor Cs\'anyi\footnote{Present address: Department of
Engineering, University of Cambridge, Trumpington Street, CB2 1PZ,
United Kingdom}}
\affiliation{Theory of Condensed Matter Group, Cavendish Laboratory,
University of Cambridge, J.~J.~Thomson Avenue, Cambridge CB3 0HE,
United Kingdom}
\author{Chris J. Pickard}
\affiliation{School of Physics and Astronomy, University of St Andrews,
North Haugh, St Andrews KY16\ 9SS, United Kingdom}
\author{B. D. Simons}
\affiliation{Theory of Condensed Matter Group, Cavendish Laboratory,
University of Cambridge, J.~J.~Thomson Avenue, Cambridge CB3 0HE,
United Kingdom}
\author{R. J. Needs}
\affiliation{Theory of Condensed Matter Group, Cavendish Laboratory,
University of Cambridge, J.~J.~Thomson Avenue, Cambridge CB3 0HE,
United Kingdom}

\begin{abstract}
Motivated by recent experimental work, we use first-principles density
functional theory methods to conduct an extensive search for low
enthalpy structures of C$_6$Ca under pressure.  As well as a range of
buckled structures, which are energetically competitive over an
intermediate range of pressures, we show that the high pressure system
($\gtrsim 18$~GPa) is unstable towards the formation of a novel class
of layered structures, with the most stable compound involving carbon
sheets containing five- and eight-membered rings. As well as
discussing the energetics of the different classes of low enthalpy
structures,
we comment on the electronic structure of the high pressure 
compound and its implications for superconductivity. 
\end{abstract}



\maketitle

Although the history of graphite intercalation compounds (GICs) dates
back more than a century, they first became prominent in the early
'60s when superconductivity was discovered in some alkali metal
GICs~\cite{Dresselhaus}. Interest in these compounds has been
reignited by the recent discovery that at least two compounds, C$_6$Ca
and C$_6$Yb,
have superconducting transition temperatures that, although modest by
the standard of cuprate compounds, are an order of magnitude higher
than those found previously for
GICs~\cite{exp_natphys,exp_emery_herold}.  Soon after this discovery,
electronic structure calculations revealed that a three-dimensional
band, originating from the intrinsic {\em interlayer} band of graphite
and a metal ion band of the intercalant,
plays a crucial role in facilitating
superconductivity in GICs~\cite{tcm_natphys} (see also
Ref.~\cite{jishi}).  Subsequently, first-principles calculations for
C$_6$Ca and C$_6$Yb~\cite{FrancescoMatteo,Mazin} provided evidence
that the out-of-plane phonons of the graphene sheets and the in-plane
phonons of the metal atom couple to the electrons in this
three-dimensional band and provide a basis for understanding the
superconductivity within a conventional BCS framework. A brief review
of the work to date is given in Ref.~\cite{MazinReview}.

As well as superconductivity, evidence for a reversible
pressure-driven phase transition has been reported in both C$_6$Ca and
C$_6$Yb~\cite{pressure_exp1, pressure_exp2}. As the pressure is
increased from ambient, the superconducting transition temperature
increases markedly, rising in C$_6$Ca from 11.5~K to 15~K at 7~GPa
above which it drops abruptly to 5~K. Similar behaviour is reported in
C$_6$Yb with a transition at ca. $2.2$~GPa.
Since the transition is reversible, and calculations have shown
softening of a phonon mode with increasing pressure, it was
conjectured that a structural phase transition takes
place~\cite{phonon_softening1,phonon_softening2}.
The reversible nature of the observed transitions suggest that they involve
small atomic displacements~\footnote{Note that such behaviour seems
incompatible with the development of staging structures.}.  However,
the various observations motivate an investigation of the wider
landscape of structures of GICs under pressure, including the
possibility of large-scale rearrangements of the atoms.  In this
letter we report a study of possible structures of C$_6$Ca under
pressure, finding that large-scale atomic rearrangements are favoured
at quite low pressures. We find that the behaviour of the C$_6$Yb
shares a similar phenomenology.

Before discussing the methodology, let us first summarise the main
findings. The results below reveal that the low enthalpy structures of
C$_6$Ca (and other GICs) can be broadly classified into different
classes.  While the planar structure of C$_6$Ca is stable at ambient
pressure, structures involving a buckling of the graphene sheets
becomes energetically competitive over a range of intermediate
pressures. At a comparatively low pressure of ca. $18$~GPa, however,
we find that the familiar honeycomb lattice structure of the graphene
layers becomes unstable towards a rearrangement involving large- and
small-diameter carbon rings, with the former accommodating the metal
ion intercalate. With such an arrangement, the volume may be reduced
without greatly increasing the internal energy, which results in a low
enthalpy.

To arrive at these conclusions, we have have carried out an extensive
search within the ``space of possible structures'' of C$_6$Ca by
relaxing a large number of random structures at a constant pressure of
$15$~GPa. (The general methodology parallels that described in
Ref.~\cite{HighPSilane}.) This approach allows a search of the
structure space with unbiased initial conditions.  The calculations
were performed within density functional theory~\cite{dft} using the
{\sc castep} package~\cite{castep}, with a plane wave basis set and
ultrasoft pseudopotentials. To remain consistent with our previous
work on the electronic structure of C$_6$Ca~\cite{tcm_natphys}, we
chose to use the local density approximation~\cite{LDA} for the
exchange-correlation functional and used the pseudopotentials that
come with the {\sc castep} package, dividing the electrons into 2 core
electrons and 4 valence electrons for C, 10 core and 10 valence for
Ca, and 42 core and 28 valence for Yb. We tested the relative
stability of the most stable structures with a generalised gradient
approximation~\cite{PBE} and found no significant changes.  The
initial conditions for the geometry optimisations consisted of a unit
cell containing one formula unit (7 atoms) with random atomic
positions and random lattice vectors, the latter bounded so that the
unit cell volume was within a factor of two of the ambient value. For
the structure-space search we used a medium quality of plane wave
cut-off energy (270~eV) and Brillouin zone integration
(0.07~\AA$^{-1}$), so that the optimisations were fast, thus enabling
a larger number of runs to be carried out using the available
computing resources (32 processor Xeon cluster). The search is
believed to be nearly exhaustive over the range of 7-atom cells
because the low enthalpy structures came up repeatedly during the
search.  The {\em Pmma} structure with a 14-atom unit cell was built
by hand using the intuition gained from the results of the random
search.

\begin{figure}
\includegraphics[width=3.2in]{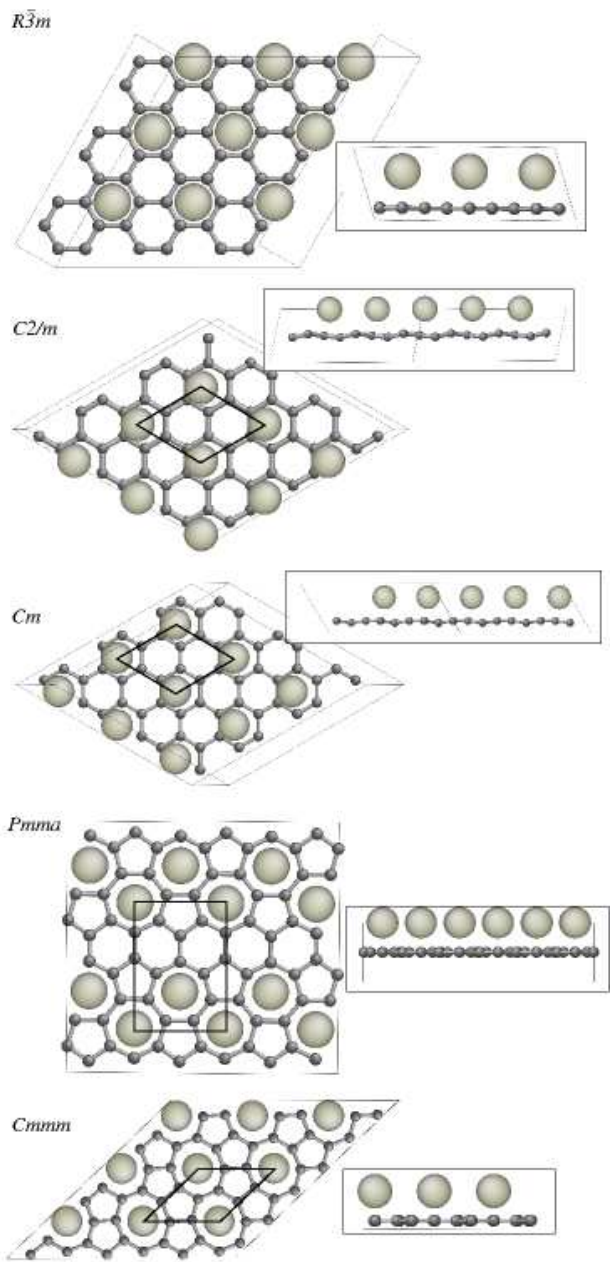}
\caption{\label{fig:structures} Low enthalpy structures of C$_6$Ca
  found during the structure space search.  The top panel shows the
  familiar $R\overline{3}m$ structure, which is stable under ambient
  conditions. The Ca ions (large) are situated between the layers,
  occupying sites above the centres of the hexagonal rings in an
  $\alpha\alpha$ stacking arrangement. The following two panels show
  lower symmetry variants, with {\em C2/m} and {\em Cm} structures,
  where in each case the graphene sheets are buckled (as depicted in
  the side views shown as insets) and the Ca ions are
  rearranged. While all of these buckled structures are energetically
  competitive, it is not possible to conclude from our calculations
  whether any become globally stable at intermediate pressures.  The
  bottom panels show structures in which the hexagonal carbon rings of
  the graphene sheets are transformed, by a sequence of Stone-Wales
  bond rotations, into five-, six- and seven-membered ({\em Pmma}) or five-
  and eight-membered rings ({\em Cmmm}). At high enough pressure, the
  latter structure is expected to become globally stable.}
\label{fig:struct}
\end{figure}

As mentioned above, the search produced a number of novel structures
which can, broadly, be grouped into two families. The first consists
of the well-known rhombohedral or trigonal $R\overline{3}m$ structure
which is the stable low-pressure phase of C$_6$Ca, and its lower
symmetry variants in which the bonding topology of the graphene sheets
remains intact, but the sheets are buckled (see
Fig.~\ref{fig:struct}). In these buckled configurations, the Ca ions
occupy the troughs between the buckled sheets, which results in a
small reduction in overall volume.  The principal difference between
the {\em Cm} and the {\em C2/m} structures is that, in the former, the
metal sheets are stacked in an $\alpha\beta\gamma$ sequence, just as
in the $R\overline{3}m$ structure, whereas the latter has an
$\alpha\alpha$ stacking.  Structures consisting of nano-porous carbon
framework filled with Ca atoms were also found, but they
were higher in enthalpy at these pressures.

Despite the buckling of the sheets, the integrity of the hexagonal
carbon ring structure in this family of compounds is maintained. In
the second family of low enthalpy compounds, the carbon and Ca atoms
maintain a planar geometry, but the hexagonal ring structure is
replaced by a network of different sized rings. The {\em Pmma}
structure has five-, six- and seven-membered rings, whereas the {\em
Cmmm} structure has only five- and eight-membered rings. A useful way
to think about these bonding topologies is that they result from one
({\em Pmma}) or two ({\em Cmmm}) Stone-Wales (SW) bond
rotations~\cite{Stone}, starting from the regular hexagonal graphene
network. The last two structures are reminiscent of those of a class
of ternary boride compounds, whose prototype is YCrB$_4$. To our
knowledge, its structure, consisting of metal ion layers and boron
sheets with five- and seven-membered, was first proposed by
Kuz'ma~\cite{Kuzma}. Several other members of the class have since
been identified~\cite{Fisk,Yvon}.

\begin{table}
\begin{tabular}{clllllll}
Space group     & \multicolumn{3}{c}{Lattice parameters}           & \multicolumn{4}{c}{Atomic coordinates} \\
                & \multicolumn{3}{c}{(\AA, $^{\circ}$)}             & \multicolumn{4}{c}{(fractional)}       \\
$Cmmm$          & $a$=9.07       & $b$=3.66       & $c$=3.54       & C1 & 0.079 & 0.000 & 0.000          \\
                & $\alpha$=90 & $\beta$=90  & $\gamma$=90    & C2 & 0.175 & 0.308 & 0.000          \\
                &                &                &                & Ca & 0.500 & 0.000 & 0.500         \\\\
$R\overline{3}m$& $a$=4.24       & $b$=4.24       & $c$=12.28      & C  & 0.000 & 0.333 & 0.167          \\
                & $\alpha$=90 & $\beta$=90  & $\gamma$=120   & Ca & 0.000 & 0.000 & 0.000          \\\\
$Pmma$          & $a$=4.82       & $b$=3.74       & $c$=6.7        & C1 & 0.250 & 0.500 & 0.117          \\
                & $\alpha$=90 & $\beta$=90  & $\gamma$=90    & C2 & 0.514 & 0.500 & 0.210          \\
                &                &                &                & C3 & 0.750 & 0.500 & 0.091         \\
                &                &                &                & C4 & 0.596 & 0.500 & 0.417         \\
                &                &                &                & Ca & 0.250 & 0.000 & 0.350         \\\\
$C2/m$          & $a$=7.30       & $b$=4.23       & $c$=4.15       & C1 & 0.000 & 0.334 & 0.500          \\
                & $\alpha$=90 & $\beta$=102 & $\gamma$=90   & C2 & 0.171 & 0.168 & 0.532          \\
                &                &                &                & Ca & 0.000 & 0.000 & 0.000         \\\\
$Cm$            & $a$=7.32       & $b$=4.24       & $c$=6.4        & C1 & 0.246 & 0.166 & 0.513          \\
                & $\alpha$=90 & $\beta$=105 & $\gamma$=90   & C2 & 0.054 & 0.332 & 0.474          \\
                &                &                &                & C3 & 0.413 & 0.333 & 0.512         \\
                &                &                &                & Ca & 0.531 & 0.500 & 1.000         \\\\
\end{tabular}
\caption{\label{table:structures} Details of C$_6$Ca structures at
15~GPa, calculated at the higher level of precision.}
\end{table}

\begin{figure}
\includegraphics[width=3.2in]{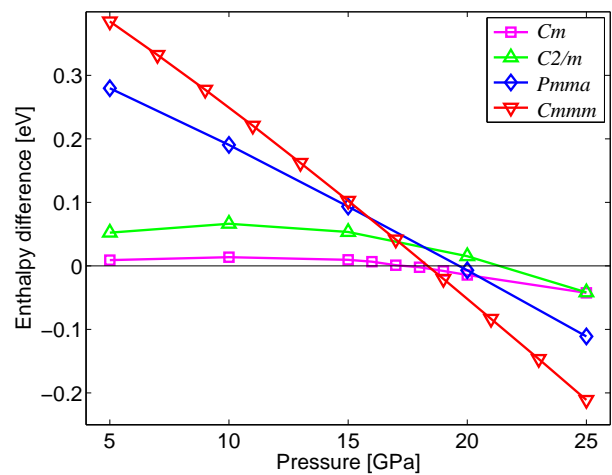}
\caption{\label{fig:enthalpy} The enthalpy differences per C$_6$Ca
  formula unit of our new structures, referenced to the
  $R\overline{3}m$ structure.}
\end{figure}

To further increase the accuracy of the procedure, we chose promising
representative structures from the different families and refined them
using a higher plane wave cut-off energy (480~eV) and Brillouin Zone
sampling (0.05~\AA$^{-1}$). The converged lowest enthalpy structures
obtained are defined in Table~\ref{table:structures}.
Figure~\ref{fig:enthalpy} shows the enthalpies of the new structures,
referenced to the $R\overline{3}m$ structure, as a function of
external pressure.  The buckled {\em Cm} structure is slightly less
stable than the $R\overline{3}m$ structure at pressures up to about
18~GPa, but at higher pressures it becomes more favourable.  In fact
the $R\overline{3}m$ and buckled structures are always close in
enthalpy, and their relative stability cannot be conclusively asserted
because of inherent systematic errors in the local density
approximation (commonly estimated as 0.05 eV/atom for total energy
differences).  The {\em Pmma} and {\em Cmmm} structures show a steep
decline in relative enthalpy, indicating that they would be favoured
at high pressure.  Considering the geometry of these structures, it
can be seen that the energy cost of the SW bond rotations is offset by
a significant reduction in volume (up to 20\%), as the metal ions are
accommodated within the larger rings.  The enthalpy-pressure curve of
{\em Pmma} has a lower slope than that of {\em Cmmm} because it
involves one SW bond rotation rather than two, so that the rings in
which the Ca atoms sit are smaller and the volume reduction is not as
large.

\begin{figure}
\includegraphics[width=3.2in]{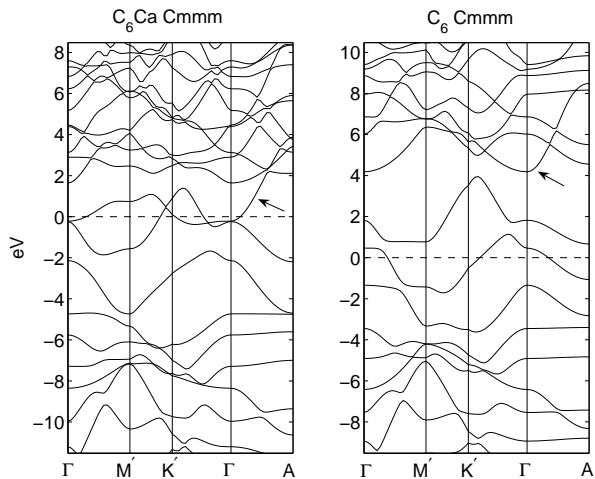}
\caption{\label{fig:bs} The band structure of the $Cmmm$ structure
  (left), and the corresponding ``empty'' structure with the metal
  ions removed (right).  The two panels have been aligned vertically
  using the bottom of the sigma bands (not shown), the respective
  Fermi levels are indicated by dashed lines.  Since the space group
  of this structure is almost hexagonal, we have used the
  corresponding notation for the points of high symmetry in the
  Brillouin Zone and indicated the small discrepancy by using primed
  letters.  Note the increase in the Fermi level upon intercalation,
  accompanied by the lowering and subsequent occupation of the
  dispersive interlayer band (marked by an arrow). }
\end{figure}

\begin{figure}
\includegraphics[width=3.2in]{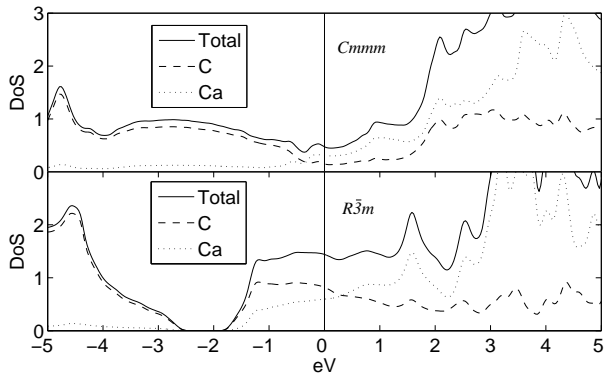}
\caption{\label{fig:dos} Density of states, resolved into
  contributions on the C and Ca atoms, 
  corresponding to the $Cmmm$ (top) and
  $R\overline{3}m$ (bottom) structures, calculated at 15~GPa.}
\end{figure}

Figure~\ref{fig:bs} shows the band structure corresponding to the {\em
Cmmm} structure. Comparing it to that of the ``empty'' carbon skeleton
(after removal of the metal ions), it can be seen that, upon
intercalation, a new dispersive band becomes occupied in much the same
way as in the parent $R\overline{3}m$ structure~\cite{tcm_natphys}.
However, the density of states (DoS, shown in Figure~\ref{fig:dos}) at
the Fermi level is significantly lower (on both the Ca and C atoms) in
the {\em Cmmm} structure, suggesting that the conditions for
superconductivity are likely to be less favourable.  In addition, the
general shape of the DoS looks much less like that of graphite as
compared with the $R\overline{3}m$ structure.

Taken together, these findings present a coherent picture of the
behaviour of C$_6$Ca under pressure. At low pressures, the graphene
sheets remain flat, leading to the familiar $R\overline{3}m$
structure. However, this parent structure is but one member of a
larger family of compounds in which the integrity of the hexagonal
rings is maintained, while the graphene sheets become buckled to
accommodate Ca atoms within the troughs. The troughs allow more space
for the Ca atoms, so these structures are favoured over
$R\overline{3}m$ at higher pressures, although the gain is limited.
At still higher pressures, another more drastic solution emerges in
which the bonding topology of the carbon atoms is disrupted by SW bond
rotations within the graphene layer. A single SW bond rotation results
in five-, six- and seven-membered rings, as in the {\em Pmma}
structure, and two SW bond rotations result in five- and
eight-membered rings as in the {\em Cmmm} structure.  The SW mechanism
results in large-diameter carbon rings, within which the Ca atoms sit,
giving a substantial volume reduction, and these structures become
stable at pressures above about 18~GPa.  The accompanying phase
transition is expected to be strongly first order, and the electronic
structure of the large-diameter-ring structures is strongly modified
from that of $R\overline{3}m$.  In summary, and recalling the
experimental evidence for a phase transition, we conclude that our
simulations strongly predict a transition to the structure like {\em
Pmma} or {\em Cmmm} at sufficiently high pressures, but due to the
intrinsic errors associated with DFT, it is uncertain whether the
observed transition is this one. It is possible that one of the
buckled structures becomes the most stable one for an intermediate
range of pressure, in which case we predict a {\em second} phase
transition at higher pressure.

Further calculations show that the bonding between the metal and
carbon atoms is stronger than might be expected, and all the
structures considered here are stable against phase separation into
diamond/graphite and pure {\em fcc}/{\em bcc} Ca.  We have also studied the
stability of above phases of C$_6$Yb under pressure, and again we find that
structures with Yb atoms sitting within larger membered rings are
stable at pressures above about 18~GPa.  We conjecture that the
occurrence of such rings accommodating the intercalate atoms might be
a general features of graphite intercalation compounds under high pressures.

\begin{acknowledgments}
CJP was supported by an EPSRC Advanced Research Fellowship. We are
grateful for G. Lonzarich for bringing to our attention the references
on the ternary borides.
\end{acknowledgments}

\end{document}